\documentclass[11pt,preprint]{aastex}
\pdfoutput=1
\usepackage{graphicx}
\let\oldfootsep=\footnotesep
\newcommand\ltsima{$\; \buildrel <\over\sim \;$}
\newcommand\simlt{\lower.5ex\hbox{\ltsima}}
\newcommand\gtsima{$\; \buildrel >\over\sim \;$}
\newcommand\simgt{\lower.5ex\hbox{\gtsima}}

\newcommand\msun {M_\odot}

\newcommand\pac{Paczy{\'n}ski }
\shorttitle{}
\shortauthors{Bennett et al}

\begin{document}

%% LaTeX will automatically break titles if they run longer than
%% one line. However, you may use \\ to force a line break if
%% you desire.

\title{MOA Data Reveal a New Mass, Distance, and Relative Proper Motion for Planetary System
OGLE-2015-BLG-0954L}

%% Use \author, \affil, and the \and command to format
%% author and affiliation information.
%% Note that \email has replaced the old \authoremail command
%% from AASTeX v4.0. You can use \email to mark an email address
%% anywhere in the paper, not just in the front matter.
%% As in the title, you can use \\ to force line breaks.

\author{D.P.~Bennett\altaffilmark{1,2},
I.A.~Bond\altaffilmark{3}, \\
and \\
F.~Abe$^{4}$, 
Y.~Asakura$^{4}$,
R.~Barry$^{1}$,
A.~Bhattacharya\altaffilmark{1,2},
M.~Donachie$^{5}$,
P.~Evans$^{5}$,
A.~Fukui$^{6}$, 
Y.~Hirao$^{7}$, 
Y.~Itow$^{4}$,  
N.~Koshimoto$^{7}$,
M.C.A.~Li$^{5}$,
C.H.~Ling$^{3}$, 
K.~Masuda$^{4}$,  
Y.~Matsubara$^{4}$, 
%T.~Matsuo$^{7}$, 
Y.~Muraki$^{4}$, 
M.~Nagakane$^{7}$,
K.~Ohnishi$^{8}$, 
C.~Ranc$^{1}$,
N.J.~Rattenbury$^{5}$, 
To.~Saito$^{9}$,
A.~Sharan$^{5}$,
%H.~Shibai$^{7}$,
D.J.~Sullivan$^{10}$, 
T.~Sumi$^{7}$,
D.~Suzuki$^{1,11}$,
P.J.~Tristram$^{12}$,
T.~Yamada$^{13}$,
T.~Yamada$^{7}$, and
A. Yonehara$^{13}$ \\ (The MOA Collaboration)\\
 } 
              
%% Mark off your abstract in the ``abstract'' environment. In the manuscript
%% style, abstract will output a Received/Accepted line after the
%% title and affiliation information. No date will appear since the author
%% does not have this information. The dates will be filled in by the
%% editorial office after submission.
%% Keywords should appear after the \end{abstract} command. The uncommented
%% example has been keyed in ApJ style. See the instructions to authors
%% for the journal to which you are submitting your paper to determine
%% what keyword punctuation is appropriate.

\affil{$^{1}$Code 667, NASA Goddard Space Flight Center, Greenbelt, MD 20771, USA;    \\ Email: {\tt david.bennett@nasa.gov}}
\affil{$^{2}$Deptartment of Physics,
    University of Notre Dame, Notre Dame, IN 46556, USA; }
\affil{$^{3}$Institute of Natural and Mathematical Sciences, Massey University, Auckland 0745, New Zealand}
\affil{$^{4}$Institute for Space-Earth Environmental Research, Nagoya University, Nagoya 464-8601, Japan}
\affil{$^{5}$Department of Physics, University of Auckland, Private Bag 92019, Auckland, New Zealand}
\affil{$^{6}$Okayama Astrophysical Observatory, National Astronomical Observatory of Japan, 3037-5 Honjo, Kamogata, Asakuchi, Okayama 719-0232, Japan}
\affil{$^{7}$Department of Earth and Space Science, Graduate School of Science, Osaka University, Toyonaka, Osaka 560-0043, Japan}
\affil{$^{8}$Nagano National College of Technology, Nagano 381-8550, Japan}
\affil{$^{9}$Tokyo Metropolitan College of Aeronautics, Tokyo 116-8523, Japan}
\affil{$^{10}$School of Chemical and Physical Sciences, Victoria University, Wellington, New Zealand}
\affil{$^{11}$Institute of Space and Astronautical Science, Japan Aerospace Exploration Agency, Kanagawa 252-5210, Japan}
\affil{$^{12}$University of Canterbury Mt.\ John Observatory, P.O. Box 56, Lake Tekapo 8770, New Zealand}
\affil{$^{13}$Department of Physics, Faculty of Science, Kyoto Sangyo University, 603-8555 Kyoto, Japan}
%\affil{$^{M}$MOA Collaboration}

%\clearpage

%% From the front matter, we move on to the body of the paper.
%% In the first two sections, notice the use of the natbib \citep
%% and \citet commands to identify citations.  The citations are
%% tied to the reference list via symbolic KEYs. The KEY corresponds
%% to the KEY in the \bibitem in the reference list below. We have
%% chosen the first three characters of the first author's name plus
%% the last two numeral of the year of publication as our KEY for
%% each reference.

\begin{abstract}
We present the MOA Collaboration light curve data for planetary microlensing event OGLE-2015-BLG-0954,
which was previously announced in a paper by the KMTNet and OGLE Collaborations.
The MOA data cover the caustic exit, which was not covered by the KMTNet or
OGLE data, and they provide a more reliable measurement of the finite source
effect. The MOA data also provide a new source color measurement that
reveals a lens-source relative proper motion of $\mu_{\rm rel} = 11.8\pm 0.8\,$mas/yr,
which compares to the value of $\mu_{\rm rel} = 18.4\pm 1.7\,$mas/yr reported in
the KMTNet-OGLE paper. This new MOA value for $\mu_{\rm rel}$ has an {\it a priori} 
probability that is a factor of $\simgt 100$ times larger than the previous value,
and it does not require a lens system distance of $D_L < 1\,$kpc. Based on the 
corrected source color, we find that the lens system consists of a planet of mass
$3.4^{+3.7}_{-1.6} M_{\rm Jup}$ orbiting a $0.30^{+0.34}_{-0.14}\msun$
star at an orbital separation of $2.1^{+2.2}_{-1.0}\,$AU and a distance of
$1.2^{+1.1}_{-0.5}\,$kpc.
\end{abstract}

%\clearpage

\section{Introduction}
\label{sec-intro}
Gravitational microlensing has a unique niche among methods for studying
exoplanet systems \citep{bennett_rev,gaudi_araa} because of its sensitivity to planets extending down to
low masses \citep{bennett96} beyond the snow line \citep{mao91,gouldloeb92},
where planet formation is thought to be the most 
efficient \citep{ida05,lecar_snowline,kennedy-searth,kennedy_snowline,thommes08},
according to the leading core accretion planet formation theory \citep{lissauer_araa,pollack96}.
Statistical analyses of exoplanetary microlensing samples have indicated that
planets with roughly Neptune masses are more common than Jupiters
\citep{sumi10,gould10,cassan12,shvartzvald16}, and the recent analysis of a larger sample
by the MOA Collaboration \citep{suzuki16} has indicated a break and likely peak in the exoplanet
mass ratio function at a mass ratio of $q \sim 10^{-4}$. This is broadly consistent
with the predictions of the core accretion theory \citep{laughlin04}.

Thus far, the statistical analyses of the exoplanets found by microlensing have
focused on the planetary parameters that are most easily measured in microlensing
events, the mass ratio, $q$, and the separation, $s$, in Einstein radius units. Fortunately,
it is possible to measure additional parameters that can constrain the lens system 
mass, $M_L$, and distance by making use of the following equations,
\begin{equation}
M_L = {c^2\over 4G} \theta_E^2 {D_S D_L\over D_S - D_L} 
= {c^2\over 4G} {{\rm AU}\over \pi_E^2} {D_S - D_L\over D_S D_L}
= {\theta_E c^2 {\rm AU}\over 4G \pi_E} \ ,
\label{eq-m}
\end{equation}
where $D_L$ and $D_S$ and the lens and source distances, $\theta_E$ is the angular Einstein 
radius, and $\pi_E$ is the microlensing parallax \citep{gould-par1,macho-par1}.
For most planetary microlensing events, 
$\theta_E$ can be directly determined from finite source effects, which provides a measurement
of the source radius crossing time, $t_*$. When $t_*$ is known, then $\theta_E$ is determined
from the expression $\theta_E = \theta_* t_E/t_*$, where $t_E$ is the Einstein radius crossing
time and $\theta_*$ is the angular source size. Both $t_E$ and $\theta_*$ are normally determined
from the model of the microlensing light curve, but $\theta_*$ generally requires a measurement
of both the source brightness and color \citep{kervella_dwarf,boyajian14}. Measurement of 
$\theta_*$ and $t_*$ also yield the lens-source relative proper motion, $\mu_{\rm rel} = \theta_*/t_*$,
which can sometimes be used to constrain the lens mass and distance based on kinematic arguments
that employ models of the stellar population of the Milky Way.

The source radius crossing time is determined for most planetary microlensing events, so
that we can determine $\theta_E$ and use the first expression of equation~\ref{eq-m} to 
constrain the lens mass and distance. For a subset of events, $\pi_E$ can be
measured due to the orbital motion of the Earth \citep{gaudi-ogle109,bennett-ogle109,bennett08,muraki11}
or observations from a satellite in a Heliocentric orbit \citep{street16}. When combined 
with a measurement of $\theta_E$, this gives the direct measurement of the lens mass given
by the last expression of equation~\ref{eq-m}.
The lens mass can also be determined with a combination of a $\theta_E$ measurement
and a measurement of the host star brightness in one or more passbands
\citep{bennett06,bennett07,bennett15,batista15,fukui15,skowron15}. Alternatively, for events without $\theta_E$
measurements, it is possible to determine the lens mass by combining measurements
of $\pi_E$ and the host star brightness \citep{kosh17_mb120950}. For complicated systems,
like the OGLE-2007-BLG-349L circumbinary planetary system
\citep{bennett16}, it is necessary to measure
$\theta_E$, $\pi_E$, and the lens system brightness to resolve all the degeneracies in the
interpretation. However, for simpler systems, the measurements of $\theta_E$, $\pi_E$, and 
the lens brightness provides redundancy that confirms that these methods are giving
reliable results \citep{gaudi-ogle109,bennett-ogle109,beaulieu16}.

The next step in the statistical analysis of wide orbit sample exoplanetary
microlens systems will be to include the constraints from measurements of 
$\theta_E$, $\pi_E$, and the host star brightness for large statistical samples,
such as that of \citet{suzuki16}. This will enable us to expand our analysis
beyond the mass ratio and separation in Einstein radius units to determine
the exoplanet mass function as a function of the host star mass, as
well as a function of Galactocentric distance. This analysis will include
not only those planetary systems with a full determination of their mass
and distance, but also events with only a measurement of a single additional
parameter, either $\theta_E$ or $\pi_E$. These partial constraints can
be useful in a Bayesian statistical analysis.

If we are going to include these additional constraints in statistical analyses of
exoplanet properties, it is important to ensure that the measurements used are
largely free of systematic errors. This is relatively straight forward to avoid mistakes
in finding the basic microlensing parameters \citep{gould_mb10523}. There has
been one recent modeling mistake \citep{udalski_ob130723,han_ob130723}, but in fact,
this was essentially identical to a much earlier modeling error 
\citep{mps-97blg41,albrow-97blg41,jung13} and not a sign of a fundamental problem
with modeling methods. However, these basic modeling errors are not the only
errors that concern us. Recent analyses have shown that it is quite possible to mistakenly attribute
excess flux seen in the vicinity of the microlensed source star to the lens star.
\citet{kosh17_mb227} have used a Bayesian analysis to show that
it is often possible for excess flux from binary companions to the lens or source stars or 
even ambient stars to be confused for the flux of the lens stars, although this effect is
much smaller if $\sim 1\msun$ stars have a much higher planet hosting probability
than low-mass stars.
\citet{aparna17} have recently shown that the excess flux detected 
for the planetary microlensing event MOA-2008-BLG-310
\citep{janczak10} does not have a relative proper motion consistent with the lens
star. Most likely, the flux is due to an ambient star, unrelated to the microlensing event.
In contrast, the excess flux seen in 4 passbands ($BVIH$) for planetary microlensing
event OGLE-2005-BLG-169 has been confirmed to have the 
$\mu_{\rm rel}$ value predicted by the microlensing light curve \citep{bennett15,batista15}.

There is statistical evidence of errors in the lens distance estimates for 31 published
exoplanets. \citet{penny16} find an excess of planetary lens systems located close to the
position of the Sun, with $D_L \simlt 1\,$kpc. This excess is likely to be due to errors
in measurement of $\pi_E$ or $\theta_E$, as errors in both parameters tend to produce
events with anomalously close distances. Microlensing parallax, $\pi_E$, is quite 
difficult to measure for distant lenses, particularly those that reside in the Galactic bulge.
So, a problem with the light curve photometry or modeling might result in a spuriously
large $\pi_E$ value, which would imply a nearby lens. In fact, this is what happened 
with OGLE-2013-BLG-0723. The original model \citep{udalski_ob130723}, which included 
a planet, had an anomalously large $\pi_E$ value. However, the paper describing the
correct, planet-free model for this event \citep{han_ob130723} shows that this anomalously large
$\pi_E$ is the result of the incorrect model trying to account for a feature that the model
cannot fully explain. MOA-2010-BLG-328 \citep{moa328} is another example of
an event with a suspiciously large $\pi_E$ value, although in this case the authors
recognized the issue and considered the possibility that the $\pi_E$ signal could be false,
a case of source orbital motion (referred to as xallarap). Another possible example is
MOA-2007-BLG-192, which was published as possible brown dwarf plus planet event,
with a planet of only a few Earth masses \citep{bennett08}. A analysis of adaptive
optics (AO) follow-up data \citep{moa192_naco} indicated the apparent detection of the
host star near the bottom of the main sequence, but this analysis did not include 
a detailed analysis of the possibility that the excess flux may not be due to the lens tar.

MOA-2014-BLG-262 is the event with the situation most similar to the event that we
discuss in this paper \citep{bennett14}. This was a relatively short duration event with
a clear signal of a planetary mass ratio companion, but the best fit model implied
a very large relative proper motion, $\mu_{\rm rel} = 19.6 \pm 1.6\,$mas/yr. 
This seemed to imply that the lens must be nearby, but the short event duration
implied a small $\theta_E$ value. From equation~\ref{eq-m}, this implies a low mass
host of only a few Jupiter masses if the host is as close as the large relative
proper motion would suggest. So, this would imply an apparently isolated planet
with a planetary mass ratio moon. However, a more careful analysis revealed another
model with a $\chi^2$ value that was only slightly larger than the best fit, but with
a larger $t_*$ value. This implied a significantly lower relative proper motion of
$\mu_{\rm rel} = 11.6 \pm 0.9\,$mas/yr, which is consistent with a lens in the Galactic
bulge. In fact, because of the small $\theta_E$ value for this event, a bulge lens
is favored.

In this paper, we consider planetary microlensing event OGLE-2015-BLG-0954. The
planetary signal was seen data both the data taken from Chile by the Optical Gravitational Lensing
Experiment (OGLE) and Korean Microlensing Telescope Network (KMTNet). They
show that their data can only be explained by a relatively high mass ratio planet
with an unusually large relative proper motion of $\mu_{\rm rel} = 18.4 \pm 1.7\,$mas/yr
by \citet{shin16}, hereafter S16.
Since the lower proper motion solution for MOA-2014-BLG-262 is strongly favored by
its prior probability, this would be the highest relative proper motion ever seen for 
a planetary microlensing event, although a non-planetary event has an even higher
relative proper motion \citep{gould09}. S16 conclude that the lens must be
located close to us, as a distance of $D_L = 0.6\pm0.3\,$kpc.

Because of an error with the MOA alert system, MOA did not issue an alert on this
event, but in fact, MOA had good coverage of this event, including coverage of the
caustic exit in good observing conditions. This feature was not observed by OGLE or KMTNet.
We model this event with MOA data in addition to the OGLE and KMTNet data,
and we find results consistent with those of S16, except for the source
color. We find a bluer source, which implies a smaller source radius, $\theta_E$
and $\mu_{\rm rel}$.

This paper is organized as follows. In Section~\ref{sec-lc_data} we describe the
published light curve data, as well as new MOA data and its photometry.
In Section~\ref{sec-lc}, we describe our light curve modeling. 
We describe the photometric calibration and the determination
of the primary source star radius in Section~\ref{sec-radius}, and then we 
derive the lens system properties in Section~\ref{sec-lens_prop}. Our
conclusions are presented in Section~\ref{sec-conclude}.

\section{Light Curve Data and Photometry}
\label{sec-lc_data}

\begin{figure}
\epsscale{0.9}
\plotone{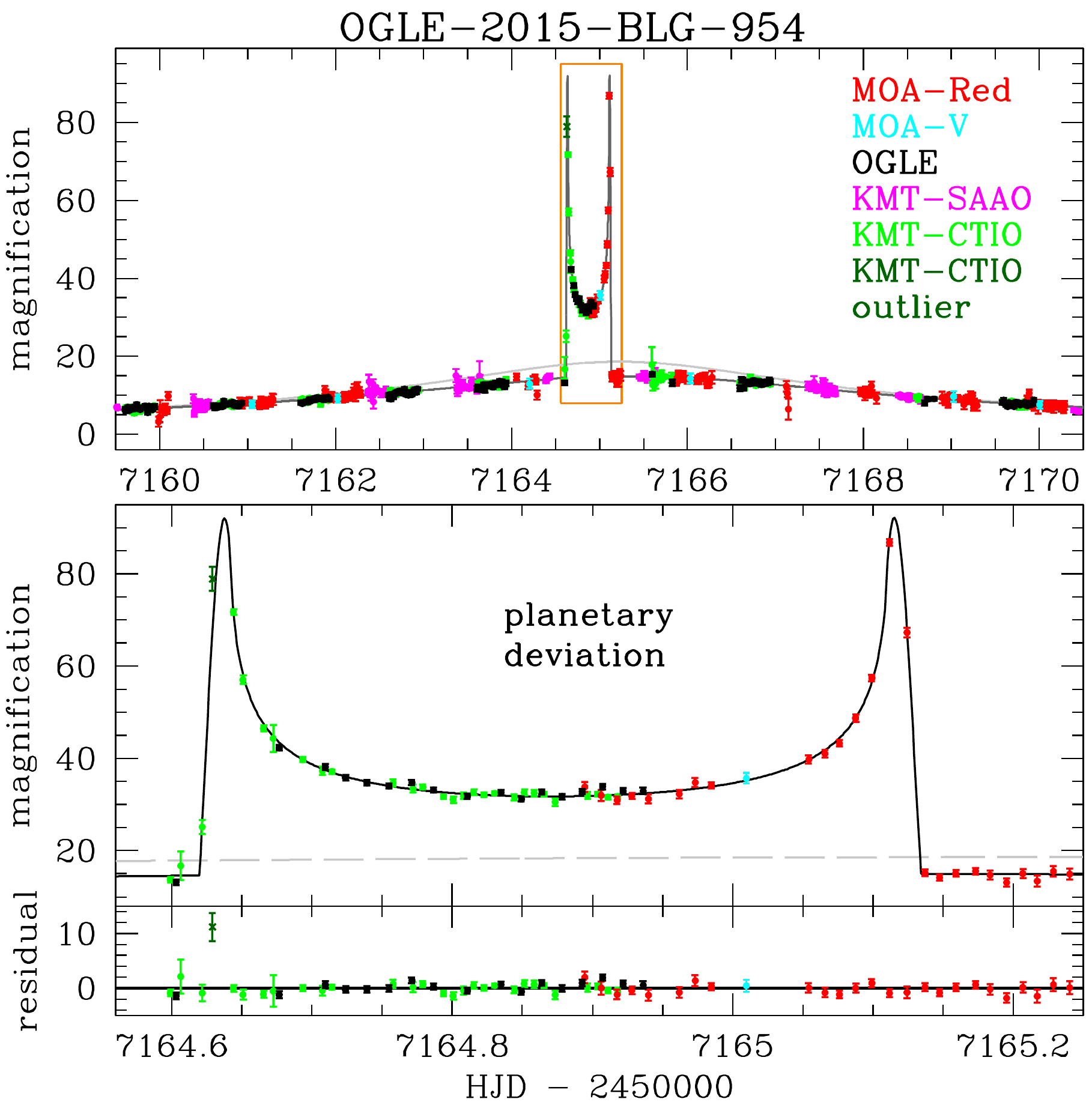}
\caption{The best binary lens model for the OGLE-2015-BLG-0954 light curve.
The $R_{\rm MOA}$ and $V_{\rm MOA}$-band data are shown in red an cyan,
while the previously published OGLE, KMT-SAAO, and KMT-CTIO $I$-band data
are shown in black, magenta, and green, respectively. The single KMT-CTIO outlier
observations near the peak of the caustic entry is shown as a dark green x.
\label{fig-lc}}
\end{figure}

The microlensing event OGLE-2015-BLG-0954 at ${\rm RA} =18$:00:44.24, 
${\rm  DEC} = -28$:39:39.2, and Galactic coordinates $(l, b) = (1.91895, -2.71366)$,
was discovered by the OGLE
collaboration \citep{ogle4} Early Warning System (EWS) \citep{ogle-ews}, and
the OGLE and KMTNet photometry was presented in the KMTNet-led 
discovery paper (S16). The observing cadence for both OGLE
and KMTNet telescope at the Cerro Tololo Inter-American Observatory (CTIO)
would normally be $\sim 20$ minutes and $\sim 15$ minutes, respectively.
However, the weather at the time of the planetary caustic
entrance was reportedly ``unstable" at CTIO and was probably poor at the OGLE telescope
at the Las Campanas Observatory (LCO), as well. As a result, the OGLE data does
not cover the caustic crossing. The KMT-CTIO coverage was better, but some
of the critical data was taken in poor observing conditions. This is probably the
reason that the KMT-CTIO observation near the top of the caustic crossing
is a $> 4$-$\sigma$ outlier as indicated in Figure~\ref{fig-lc}.

The MOA Collaboration did not identify this event with its alert system, but we
found a strong signal including coverage of the caustic exit, indicating that the event
was missed due to an alert system \citep{bond01} error, probably an oversight by the
observer. We obtained optimized photometry via a reduction of $R_{\rm MOA}$ and 
$V_{\rm MOA}$ images obtained by extracting sub-images centered on the event \citep{bond17}. 
For this analysis, we used observations from April, 2012 through August, 2016. We used
our own difference imaging implementation that incorporates a numerical kernel as described 
by \citep{bramich08} with our own modification to allow for a spatial variation of the kernel 
across the field-of-view in a similar manner to that given by \citep{alard00}. The photometry for both
the reference and difference images were measured using an analytic PSF model 
of the form used in the DoPHOT photometry code \citep{dophot}. Trends in the photometry
with seeing, airmass, and differential refraction (parameterized by the hour angle and airmass)
were removed based on the observed trends in 2012-2014 and 2016. The MOA photometry
consists of 252 observations in the $V_{\rm MOA}$ passband and 9663 $R_{\rm MOA}$ after
removing 7 4-$\sigma$ outliers from the best fit model, which were all separated by more than
a week from the planetary signal.
The MOA instrumental magnitudes were calibrated by cross referencing stars in the DoPHOT 
catalog to stars in the OGLE-III catalog which provides measurements in the standard 
Kron-Cousins $I$ and Johnson $V$ passbands \citep{ogle3-phot}. This allows us
to derive relations to calibrate the MOA photometry to standard magnitudes.

We obtained the KMTNet and OGLE $I$-band light curve data from KMTNet group, and 
we used all of the OGLE and KMT-SAAO data that was provided. The KMT-CTIO
data consisted of 3 separate light curves due to changes in the camera electronics. (This
was the first year of the KMTNet survey.) Since the long term behavior of the light curve
is well covered by MOA and OGLE, we include only the second portion of the 
KMT-CTIO light curve, which consists of 1187 observations. However, the data
point closest to the peak of the light curve is a 4-$\sigma$ outlier, so we do not
include that data point in our modeling.

\section{Light Curve Models}
\label{sec-lc}

Our light curve modeling was done using the image centered ray-shooting method
\citep{bennett96}, and we employed the initial condition grid search
method outlined in \citet{bennett-himag} to search the binary lens parameter space
for solutions. Unsurprisingly, we recovered solutions very similar to those presented
in S16. The parameters of our best fit close and wide models are shown
in Table~\ref{tab-mparams}, along with
the Markov Chain Monte Carlo (MCMC) results for solutions centered on each
best fit model. The model parameters that also apply to a single lens system are 
the Einstein radius crossing time, $t_E$, and the time, $t_0$, and distance,
$u_0$, of closest approach between the lens center-of-mass and the source star.
Binary lens models also include the mass ratio of the secondary to the primary lens,
$q$, the angle between the lens axis and the source trajectory, $\theta$, and the
separation between the lens masses, $s$. Figure~\ref{fig-lc} shows the light
curve and best fit close model, which is slightly favored over the best fit wide
model. The length parameters, $u_0$ and $s$, are normalized by the Einstein radius of this total system mass, 
$R_E = \sqrt{(4GM/c^2)D_Sx(1-x)}$, where $x = D_L/D_S$ and $D_L$ and $D_S$ are
the lens and source distances, respectively. ($G$ and $c$ are the Gravitational constant
and speed of light, as usual.) 

\begin{deluxetable}{cccccc}
\tablecaption{Model Parameters
                         \label{tab-mparams} }
\tablewidth{0pt}
\tablehead{
%% Use a footnote to explain numbering.
& & & & \multicolumn{2}{c} {MCMC averages} \\
\colhead{parameter}  & \colhead{units} &
\colhead{$s<1$} & \colhead{$s<1$} &\colhead{ $s<1$} & \colhead{$s>1$} 
}  % end header.

\startdata

$t_E$ & days & 39.603 & 39.386 & 39.5(1.2) & 39.4(1.2) \\
$t_0$ & ${\rm HJD}-2450000$ & 7165.2277 & 7165.1763 & 7165.227(9) &  7165.176(8) \\
$u_0$ & &0.053803 & 0.047302 & 0.0541(20) & 0.0473(17)  \\
$s$ & & 0.79750 & 1.35247 & 0.7977(32) &  1.3529(30) \\
$\theta$ & radians & 1.40353 & 1.40857 & 1.4039(41) & 1.4085(39)  \\
$q$ &  & 0.01031 & 0.01134 & 0.01035(26) & 0.01136(31)  \\
$t_\ast$ & days & 0.01153 & 0.01148 & 0.01152(28) & 0.01144(29)  \\
$I_S$ & & 21.053 & 21.038 & 21.047(33) & 21.037(32)  \\
$V_S-I_S$ & & 1.702 & 1.702 & 1.702(31) & 1.702(31)  \\
fit $\chi^2$ &  &16820.62 & 16821.20 &  &   \\

\enddata
\end{deluxetable}

For every passband, there are two parameters to describe the unlensed source
brightness and the combined brightness of any unlensed ``blend" stars that are
superimposed on the source. Such ``'blend" stars are quite common because
there are usually several relatively bright, main sequence stars in each $\sim 1^{\prime\prime}$
seeing disk. These stars generally are not microlensed because this requires 
lens-source alignment of $\simlt \theta_E \sim 1\,$mas.
The source and blend fluxes are treated differently from the other parameter because
the observed brightness has a linear dependence on them, so for each set of 
nonlinear parameters, we can find the source and blend fluxes that minimize the
$\chi^2$ exactly, using standard linear algebra methods \citep{rhie_98smc1}.

The source radius crossing time, $t_*$, is an important parameter because
the both $\theta_E = \theta_* t_E/t_*$ and $\mu_{\rm rel} = \theta_*/t_*$ depend on it.
Because of KMTNet data had a 4-$\sigma$ photometric outlier on the one
caustic crossing resolved in the discovery paper (S16), we had expected
a high probability that our best fit $t_*$ values would differ from the S16
values. However, our values are $t_* = 0.01152\pm 0.0028$ and $0.01144\pm 0.0029\,$days for
the close and wide models, respectively. These compare to the S16 values of
$t_* = 0.0111\pm 0.004$ and $0.112\pm 0.004\,$days for the close and wide models,
so our $t_*$ values are larger by only about 1-$\sigma$. Thus, our new $t_*$
values will not reduce the $\mu_{\rm rel}$ value reported by S16
by a significant amount.

\section{Photometric Calibration and Source Radius}
\label{sec-radius}

The measurement of the angular source radius, $\theta_*$, can have a significant
effect the lens-source relative proper motion measurement. This requires the
determination of the extinction corrected source brightness and color. The first
step is to convert our instrumental $R_{\rm MOA}$ and $V_{\rm MOA}$ magnitudes
to calibrated OGLE-III magnitudes \citep{ogle3-phot}. Using the photometry described in 
Section~\ref{sec-lc_data}, we derive
\begin{eqnarray}
& I_{\rm O3} = 28.0205\pm 0.0039+ R_{\rm MOA} - (0.2183 \pm 0.0030)\left(V_{\rm MOA}- I_{\rm MOA}\right)
\label{eq-I-cal} \\
& V_{\rm O3}-I_{\rm O3} = 0.4188 \pm 0.0054 + (1.0725\pm 0.0042) \left(V_{\rm MOA}- R_{\rm MOA}\right)
\label{eq-VmI-cal} \ ,
\end{eqnarray}
\\
where we have used comparison stars with $I_{\rm O3} \leq 16.0$ to derive the transformation.
Although the MOA-red filter is roughly equivalent to Cousins $R$ plus Cousins $I$, the
high extinction of the bulge fields reduces the bluer flux and pushes the $R_{\rm MOA}$
to be closer to Cousins $I$ rather than Cousins $R$.

\begin{figure}
\epsscale{0.9}
\plotone{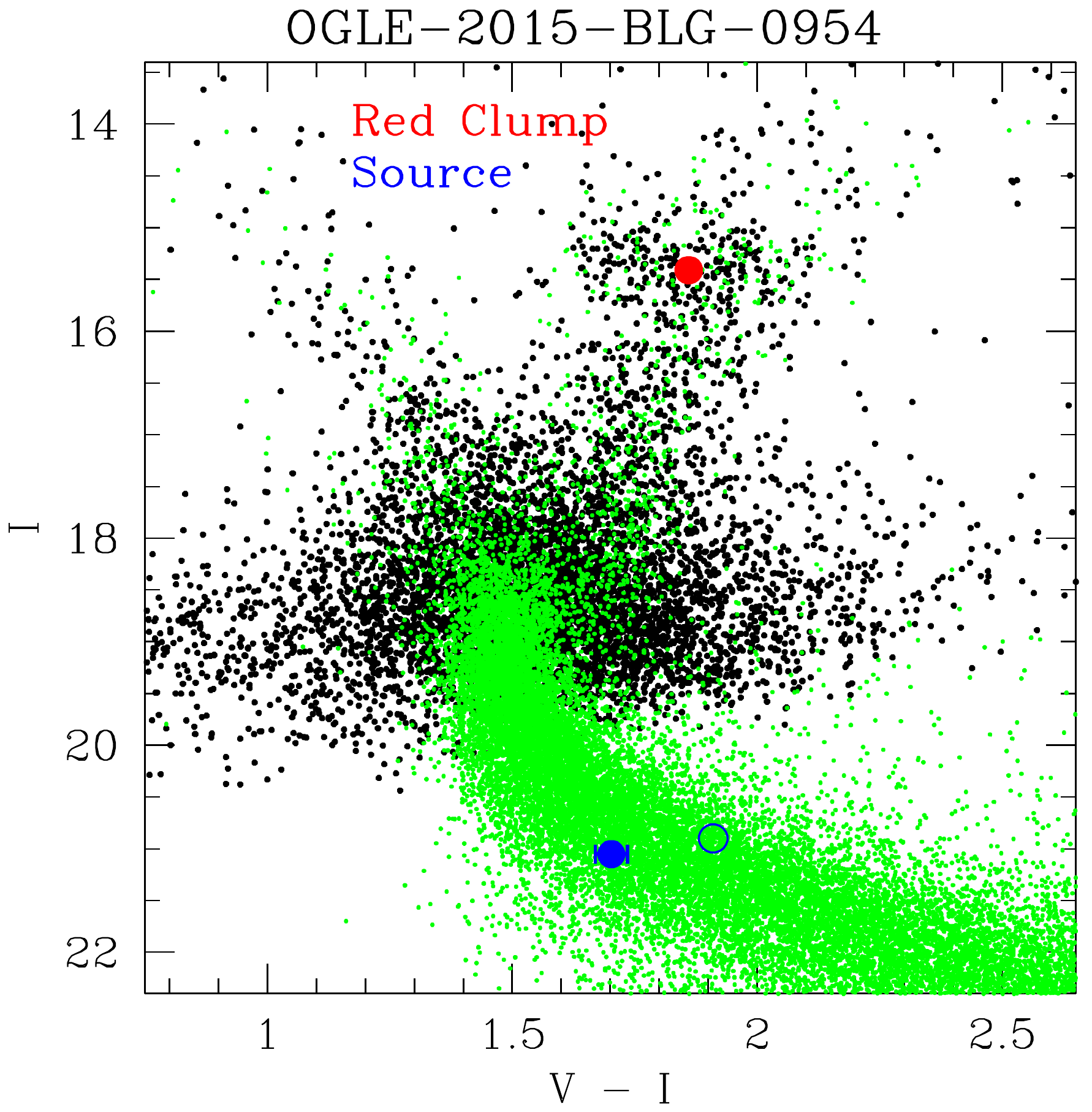}
\caption{The $(V-I,I)$ CMD of the stars in the OGLE-III catalog
\citep{ogle3-phot}
within $90^{\prime\prime}$ of OGLE-2016-BLG-0954. The red spot indicates
red clump giant centroid, and the solid blue spots indicates the source magnitudes
and color, while the open blue circle indicates the color estimate from S16.
The green spots indicate the HST CMD from \citet{holtzman98} shifted to the bulge distance
and extinction at the OGLE-2016-BLG-0954 line-of-sight.
\label{fig-cmd}}
\end{figure}

The next step is to determine the extinction. Figure~\ref{fig-cmd} shows the
color magnitude diagram (CMD) of the stars in the OGLE-III catalog within
$90^{\prime\prime}$ of OGLE-2016-BLG-0954 along with the Baade's Window
HST CMD \citep{holtzman98} (in green) shifted to the extinction and bulge distance for
the position of our target. We determine the red clump centroid to be located at
$(V-I)_{\rm rc} = 1.86$, $I_{\rm rc} = 15.41$, whereas \citet{nataf13} predict the
unextincted red clump centroid to be located at $(V-I)_{\rm rc,0} = 1.06$, $I_{\rm rc,0} = 14.375$
at this Galactic longitude. This implies $I$ and $V$-band extinctions of $A_I = 1.035$
and $A_V = 1.835$. From Table~\ref{tab-mparams}, we see that the best fit 
source magnitude and color are $I_S = 21.053$ and $(V-I)_S = 1.702$, so that the
unextincted source magnitude and color are $I_S = 20.018$ and $(V-I)_S = 0.902$.
We determine the angular source radius using
\begin{equation}
\log_{10}\left[2\theta_*/(1 {\rm mas})\right] = 0.501414 + 0.419685\,(V-I)_{s0} -0.2\,I_{s0} \ ,
\label{eq-theta_star} 
\end{equation}
which comes from the \citet{boyajian14} analysis using a restricted set of data using only 
stars with $3900<T_{\rm eff}<7000$ (Boyajian, private communication, 2014). 
This gives $\theta_* = 0.376\pm 0.022\,\mu$as for the best fit model, which is a factor of 1.49 smaller
than the S16 value of $\theta_* = 0.56$. This difference is due to the fact
that we find the source 0.14 mag fainter and 0.28 mag redder than the values
published by S16, as indicated by the open blue circle in Figure~\ref{fig-cmd}. 
Because the coefficient of the $(V-I)_{s0}$ term in
Equation~\ref{eq-theta_star} is a little more than twice as large as the coefficient of the $I_{s0}$ term,
we can see that $\sim 80$\% of this difference is due to the difference in
color measurements, while the remaining $\sim 20$\% is due to our slightly larger
$t_E$ value and fainter source.

This new $\theta_*$ implies an angular Einstein radius of $\theta_E = 1.292 \pm 0.075$, where the
error bar includes only the contribution from the $\theta_*$ uncertainty. (The contribution from 
the model parameter uncertainties will be presented in Section~\ref{sec-lens_prop}.)
This $\theta_E$ value is a factor of 1.46 smaller than the S16 value, but it is still
much larger than average. From Equation~\ref{eq-m}, we can see that it implies
a mass of $1.6 \msun$ for a source at $D_S = 8\,$kpc and a lens at $D_L = 4\,$kpc.
However, the OGLE and MOA data are not compatible with a main sequence 
star any more massive than $\sim 0.9 \msun$ main sequence star at $4\,$kpc
at the position of the source (as we discuss below in Section~\ref{sec-lens_prop}).

\subsection{Comparison of Source Color Measurements}
\label{sec-comp_color}

Since there are many more $R_{\rm MOA}$ measurements than $V_{\rm MOA}$
measurements, the uncertainty in the color measurement is determined by the uncertainty
of the source brightness in the $V_{\rm MOA}$ band.
Most of the magnified data responsible for our $V_{\rm MOA}$ band measurements is 
shown Figure~\ref{fig-lc} as the cyan colored points. There is one observation per
day in the relatively good observing conditions near the peak of this event. 
There are 7 $V_{\rm MOA}$ data points with magnification $> 7$, and these points
have a total $\chi^2 = 1.19$. This corresponds to a $\chi^2/{\rm dof} = 0.24$ if we
assume that these 7 data points control both the source and blend brightness 
parameters for the $V_{\rm MOA}$ band. This small $\chi^2$ value is correlated
with relatively small photometric error bars on these data points, which implies
that the  $V_{\rm MOA}$ band data near the light curve peak should be reliable.
There are many data points at low magnification, so we expect that neither the
high magnification points nor the low magnification points should be affected by
systematic errors due to photometric irregularities. 

It is more difficult to investigate potential photometric problems with the
OGLE and KMT-CTIO $V$-band data, because we do not have access to
the data. The OGLE $I$-band data have similar scatter to $R_{\rm MOA}$
data interior to the caustic crossing, but S16 report 
$(V-I)_{s,\rm OGLE} = 1.91 \pm 0.07$. Since this error bar is 2.3 times larger
than the $(V-I)_{s}$ derived from the MOA data, it seems unlikely that OGLE
has any $V$-band observations interior to the caustic crossing.

The situation is different for the KMT-CTIO data, as S16 report that 
one out of every 6 images are taken in the $V$-band by this telescope.
Since the KMT-CTIO $I$-band data set contains 27 data points above
magnification 30 on the night of the caustic entry, we might expect
that they also have $\sim 5$ $V$-band at magnification $> 30$.
However, S16 report a KMT-CTIO color estimate of $(V-I)_{s,\rm KMT-C} = 2.01 \pm 0.05$,
so the error bar is also larger than the $(V-I)_{s}$ derived from the MOA data,
despite the likelihood of many more $V$-band observations than MOA.
While we cannot be certain of the situation with the KMT-CTIO $V$-band data,
there are several causes for concern. First, the light curve data were reduced with
DoPHOT instead of difference imaging. This is often adequate for bright sources
in good observing conditions, but this event was quite faint, except for the
night of the caustic crossing when weather conditions at CTIO were reportedly
unstable. 

Second, the color is not determined by a fit to the light curve. 
Instead, it is determined by a linear fit to $V$ and $I$-band observations
that are approximately simultaneous. This is a common method used when
there is some uncertainty about the correct model. However, it is subject
to large systematic errors if the magnification changes significantly during
the time interval that is considered ``approximately simultaneous\rlap."
S16 use the criteria that two observations can be consider simultaneous
if they are separated by $\leq 0.05\,$days, which is the interval between
the tick marks in the lower two panels of Figure~\ref{fig-lc}. This
is a poor assumption for the night of the caustic entry. According to the
best fit model, the event brightens by a factor of $>6$ in the 
$ 7164.6 < t < 7164.65$, and then drops by factors of 1.51, 1.13, and 1.055
over the intervals $ 7164.65 < t < 7164.7$, $ 7164.7 < t < 7164.75$, and
$ 7164.75 < t < 7164.8$, respectively. It is only after $t = 7164.8$ that the
light curve variation over this 0.05 interval drops below the quoted 0.05
mag uncertainty of the color measurement, but two thirds of the
KMT-CTIO observations were taken before $t = 7164.8$ on the night of the
caustic crossing. For these reasons, we do not use the KMT-CTIO color
estimate in our calculations.

We also have the option of using the OGLE color of 
$(V-I)_{s,\rm OGLE} = 1.91 \pm 0.07$, which is at most, marginally
inconsistent with the MOA color. The weighted combination of these
measurements is $(V-I)_{s,\rm MO} = 1.736 \pm 0.026$. This is only
1.1-$\sigma$ above the MOA value of $(V-I)_{s,\rm M} = 1.702\pm 0.031$,
so the inclusion of the OGLE measurement would have little effect on our 
conclusions.

The Bayesian analysis presented in the next section can also be used to 
compare the {\it a priori} probabilities for the MOA and KMTNet-OGLE
$\mu_{\rm rel}$ values. This analysis indicates that $\mu_{\rm rel} = 11.8\,$mas/yr
is 111 times more likely than $\mu_{\rm rel} = 18.4\,$mas/yr. Because both cases
favor nearby lenses, these analyses are not dependent on the detailed structure
of the Galactic bulge model, but they do depend on the high velocity tails of
the velocity distribution for all the stellar components of the Galaxy, which 
our method is optimized for \citep{bennett14}.

\section{Lens System Properties}
\label{sec-lens_prop}

\begin{deluxetable}{cccc}
\tablecaption{Physical Parameters\label{tab-pparam}}
\tablewidth{0pt}
\tablehead{
\colhead{Parameter}  & \colhead{units} & \colhead{value} & \colhead{2-$\sigma$ range} }
\startdata 
$D_L $ & kpc & $1.2^{+1.1}_{-0.5}$ & 0.4-3.0 \\
$M_{\rm host}$ & $\msun$ & $0.30^{+0.34}_{-0.14}$ & 0.09-0.83 \\
$m_p$ & $M_{\rm Jup}$ & $3.4^{+3.7}_{-1.6}$ & 1.0-9.3 \\
$a_\perp$ & AU & $1.6^{+1.3}_{-0.7}$ & 0.5-4.5  \\
$a_{3d}$ & AU & $2.1^{+2.2}_{-1.0}$ & 0.6-9.3  \\
$\mu_{\rm rel}$ & mas/yr & $11.8\pm 0.8$ & 10.3-13.3 \\
$\theta_E$ & mas & $1.28 \pm 0.08$ & 1.12-1.44 \\
%$P$ & yr & $7.6{+7.7\atop -1.4}$ & 5.4-62 \\
%$V_L$ & mag & $22.81{+0.09\atop -0.07}$ & 22.68-23.05 \\
$I_L$ & mag & $20.61^{+0.56}_{-0.11}$ & 18.55-21.78 \\
$K_L$ & mag & $17.45^{+0.33}_{-0.53}$ & 16.27-18.12 \\
\enddata
\tablecomments{ Uncertainties are
1-$\sigma$ parameter ranges. }
\end{deluxetable}

Because we are lacking a microlensing parallax measurement and a lens brightness
measurement, we are unable to use the expressions in Equation~\ref{eq-m} to 
directly determine the lens mass and distance. Instead, we are limited to a
Bayesian analysis using the $\theta_E$ mass-distance relation in Equation~\ref{eq-m}.
This is similar to the situation for event MOA-2011-BLG-262 \citep{bennett14} because
the $\mu_{\rm rel}$ value is larger than for most microlensing events, so we use
the Galactic model employed for the \citet{bennett14} analysis because it was designed to
include the high velocity components of the galaxy, such as the thick disk and the spheroid
(or stellar halo), while also enforcing the Galactic escape velocity as an upper limit on
the velocities. The large $\mu_{\rm rel}$ value does favor nearby lenses, but for the
MOA-2011-BLG-262 event, with a slightly larger $\mu_{\rm rel}$, a lens location in the Galactic
bulge received comparable probability to nearby lens locations. 
The situation for OGLE-2015-BLG-0954 is different because 
of its much larger $t_E$ and $\theta_E$ values. The mass-distance relation from 
Equation~\ref{eq-m} implies that a bulge lens would have to be faint but also very
massive. So, it could only be 
a stellar mass black hole. We do not consider black hole hosts in our analysis,
but if we did, the probability of such a lens system would still be small because stellar
mass black holes are known to be rare compared to main sequence stars even though
their occurrence rate is not very well known.

The results or our Bayesian analysis are given in Table~\ref{tab-pparam}.
The main difference with the S16 results is that the lens system is likely to
be at a larger distance of $D_L = 1.2^{+1.1}_{-0.5}\,$kpc, compared to
$D_L = 0.6\pm 0.3$ claimed by S16. (This is derived from their equation 14.)
With our smaller $\theta_E$ and $\mu_{\rm rel}$ values, we find that this event is
consistent with a large range of host star and planet masses. This new distance
reduces the tension seen by \citet{penny16} in the number of planet discoveries
estimated to be at small distances, $D_L < 1\,$kpc. Our 68\% and 95\%
for the host mass are 0.16-$0.64\msun$ and 0.09-$0.83\msun$, respectively.
The 2-$\sigma$ upper limit on the host mass is due to the observed upper limit 
on the host star flux (assuming a main sequence host star). S16 reported
2 different mass values: an upper limit of $M_{\rm host} < 0.25\msun$ for
a main sequence host and $M_{\rm host} = 0.33\pm 0.12\msun$
for a host of any type. These are an artifact of their apparent color
measurement error, and the true range of host masses is much larger.
This larger range of host masses implies a large range of planet masses
with 68\% and 95\% confidence level ranges of 1.8-$7.1 M_{\rm Jup}$ and
0.9-$9.3 M_{\rm Jup}$. The ranges of three-dimensional separations
predicted for this event are 1.1-$3.3\,$AU and 0.6-$9.3\,$AU at 
the  68\% and 95\% confidence levels.

The $I$ and $K$ band brightness ranges for the lens (and host) star are
also given in Table~\ref{tab-pparam}. The 95\% confidence level ranges are
$18.55\leq I_L \leq 21.78$ and $16.27\leq K_L \leq 18.12$, which compare
to the measured source brightness (from Table~\ref{tab-mparams}) of
$I_S = 21.04\pm 0.03$. This implies that the lens star will be somewhere
between one half and and ten times as bright as the source in the $I$-band,
with a smaller range of brightness in the $K$-band. The {\it HST} analysis of
\citep{aparna17} implies that the lens-source separation can already
be measured with {\it HST} images, while the two stars should be resolvable
with Keck AO images by 2020 \citep{batista15}. The measurement of the
host star brightness will allow the host mass to be determined through
the combination of the mass distance relation (equation~\ref{eq-m}) and
a mass-luminosity relation \citep{bennett06}.

\section{Discussion and Conclusions}
\label{sec-conclude}

At present, all the statistical analyses of the properties of planetary 
systems found by microlensing \citep{sumi10,gould10,cassan12,shvartzvald16,suzuki16}
have characterized the distribution with the mass ratio, $q$, and projected separation, $s$, of the
discovered planets, although \citet{suzuki16} did also look at the $t_E$ dependence.
However, we now have a growing number of measurements off additional
parameters that can be used to learn more details of the exoplanet properties
beyond the snow line. For a growing number of events, we can determine masses
from microlensing parallax measurements, either from the orbital motion of the
Earth \citep{gaudi-ogle109,bennett08,bennett16,muraki11,han13,moa328,skowron15,
sumi16,kosh17_mb120950} or from observations from a satellite in a Heliocentric orbit \citep{street16}.
High angular resolution follow-up observations can allow the determination of the planet and
host star masses using the mass-distance relation given in equation~\ref{eq-m} and a mass-luminosity
relation \citep{bennett06,bennett15,batista15,dong-ogle71,fukui15,beaulieu16}. The
combined results of these $\theta_E$, $\pi_E$, and host star brightness measurements will
be to allow us to expand our statistical analysis beyond the simple mass ratio and separation
measurements to include dependence of exoplanet properties on the host star mass and 
Galactocentric distance. So, microlensing had the potential to give us a comprehensive
picture of planets that orbit beyond the snow line throughout the Galaxy.

In order to realize the potential of these additional measurements, it is important to
ensure that these additional measurements are free of errors. \citet{penny16} noted
a statistical indication of such errors in the large number of published planetary
microlensing events with lens distances of $D_L < 2\,$kpc and especially 
$D_L <1\,$kpc. The error that yielded one of these events has already been
found  \citep{udalski_ob130723,han_ob130723}, as a large spurious parallax
signal resulted from using the wrong lens model geometry. A different type of 
difficulty has been identified by \citet{aparna17} and \citet{kosh17_mb227}.
They have identified cases where excess flux on top of the source star could
be mistakenly attributed to the lens star. The solution to this difficulty is to
ensure that the lens-source relative proper motion matches the model prediction
\citep{bennett15,batista15}.

The problem that we have identified with the S16 analysis of OGLE-2015-BLG-0954
is a problem with the source color measurement. We have not seen the $V$-band
data used for the S16 paper, so we can not be sure why our results disagree with
theirs. However, the difference imaging method \citep{tom96,ala98} has been the state-of-the-art
crowded field photometry method for microlensing surveys for nearly two decades
\citep{macho-DIA1,ogle-pipeline}. For some difference imaging implementations,
it has been difficult to put the light curve photometry on the same photometric
scale as the photometry of the stars in the reference frame because different PSF
models are used in each case. However, the OGLE Collaboration solved this problem
with its pipeline back in 2003 \citep{ogle-pipeline}. \citet{bennett12} showed that it is
straight forward to modify a DoPHOT-like \citep{dophot} code to produce photometry
of difference images on the same photometry scale as the DoPHOT reference frame
photometry, and the MOA collaboration has developed the code used in this paper
for this task \citep{bond17}. We recommend that difference imaging photometry
should always be used for color measurements in the future.

The implied properties of the OGLE-2015-BLG-0954Lb planet are
not greatly effected by the change in the $\theta_E$ and $\mu_{\rm rel}$ values.
The previous, relatively tight, limits on the mass of the host star and planet
are now substantially weakened. The host star is no longer limited to be a
low mass star with $M_L < 0.25\msun$, and instead is in the range
$0.09\,\msun < M_L < 0.83\,\msun$ (at 95\% confidence). Fortunately, if the
lens is a main sequence star, it should be easily detectable with {\it HST}
or AO observations.

\acknowledgments 
D.P.B., A.B., D.S.\ and C.R.\  were supported by NASA through grants NASA-NNX13AF64G
and NNX16AN59G.
Work by C.R.\ was also supported by an appointment to the NASA Postdoctoral Program at the 
Goddard Space Flight Center, administered by USRA through a contract with NASA.
The MOA project in Japan is supported by JSPS KAKENHI grants JSPS24253004, 
JSPS26247023, JSPS23340064, JSPS15H00781, JP16H06287, and JP17H02871.

\end{document}